\documentclass[aps,pra,showpacs,superscriptaddress,floatfix]{revtex4}
\pdfoutput=1
\usepackage{graphicx}
\usepackage{color}
\usepackage{amsmath}
\usepackage{amsfonts}

\begin{document}

\title{
A quantum spin approach to histone dynamics}
\author{C. Gils}
\affiliation{Samuel Lunenfeld Research Institute, Mount Sinai Hospital, 600 University Ave, Toronto, ON M5G 1X5, Canada}
\author{J.~L. Wrana}
\affiliation{Samuel Lunenfeld Research Institute, Mount Sinai Hospital, 600 University Ave, Toronto, ON M5G 1X5, Canada}
\affiliation{Department of Molecular Genetics, University of Toronto, 1 Kings College Circle, Room 4396, Toronto, ON M5S 1A8, Canada}
\author{W.~K. Abou Salem}
\affiliation{Department of Mathematics, University of British Columbia, 1984 Mathematics Road, Vancouver, BC V6T 1Z2, Canada}
\affiliation{Department of Mathematics and Statistics, University of Saskatchewan, Saskatoon, SK S7N 5E6, Canada}
\date{\today}

\begin{abstract}

Post-translational modifications of histone proteins are an important factor in epigenetic control 
that serve to regulate transcription, depending on the particular modification states of the histone proteins.
We study the  stochastic dynamics of  histone protein states,   
   taking into account  a  feedback mechanism  where modified nucleosomes recruit enzymes  that  diffuse to
   adjacent nucleosomes.  
   We map the system  onto a quantum spin system whose dynamics is 
   generated by a non-Hermitian
   Hamiltonian. 
   Making an ansatz for the solution as a tensor product state leads to nonlinear partial differential
   equations that describe the dynamics of the system. Multiple stable histone states appear in a 
   parameter regime whose size increases with increasing number of 
  modification sites. We discuss the role of the spatial dependance, and we
 consider the effects of spatially  heterogeneous   enzymatic activity.  
Finally, we consider multistability in a model of several types of correlated post-translational modifications.
 \end{abstract}


\maketitle


\section{Introduction}

Nuclear chromosomes in eukaryotic organisms consist of the chromatin, a 
 complex wrap  that is primarily composed of DNA and histone proteins.  
 The fundamental unit of the chromatin is the nucleosome, each of which  contains two  copies of the core histones 
 H2A, H2B, H3 and H4, and approximately $150$ base pairs of DNA.
Each of the core histone proteins exhibits multiple
 amino acid residues that are subject to post-translational modifications (PTM) by 
 chemical groups such as 
phospho-, acetyl-,  methyl- or ubiquitin-groups that can be added and removed in a reversible manner.
For example, H4 has a phosphorylation site, four acetylation sites and six methylation sites.
Depending on the  particular modification state of histones, certain regions of DNA in the chromatin 
 are in an active or repressed state.
Regulation of the PTMs of histones 
 lies at the center of epigenetic control
  \cite{Allis:07,Peterson:04,Rando:09}.

A commonly observed epigenetic phenomenon is the existence of alternative  regulatory  states. 
  For example, 
    in the fission yeast Schizosaccharomyces pombe  the two mating type cassettes, mat2-P and mat3-M are
    usually in a  silenced state in which the mating type genes are not expressed. 
    When removing a portion of the silenced region and inserting a ura4+ reporter gene, the
     expression of ura4+ and the mating-type genes becomes bistable, with a
      state where ura4+ is repressed and a state where ura4+ is expressed~\cite{Grewal:96,Thon:96,Grewal:02}.
      The silenced state of ura4+ is associated with a high concentration of methylation marks 
 on lysine of histone H3 (H3K9),   while
       the active ura4+ state does not exhibit methylation of  H3K9~\cite{Hall:02}.
   Each of the two epigenetic states is preserved under cell divisions, with transitions between them occuring only
 at a very low rate.

Post-translational modifications are  regulated by various enzymes.
  In order to explain the appearance of multiple stable histone states, a 
     non-local positive  feedback mechanism has been put forward~\cite{Turner:98,Grunstein:98}: 
 A nucleosome that 
 exhibits a particular modification recruits the enzymes that catalyze this  modification. These enzymes then  move
  to adjacent nucleosomes and cause the modification to be added there, a mechanism that has indeed been observed for 
 some histone acetyltransferases, histone decacetylases and histone methyltransferases~\cite{Jacobsen:00,Owen:00,Rusche:01,Schotta:02}.
Long-range  feedback has been implemented in a stochastic simulation of a three-state model
     (unmodified state, acetylated state, methylated state) and it
     was shown to lead to robust bistability~\cite{Dodd:07}.
Nearest-neighbour feedback has been considered in deterministic descriptions of two- and three-state models~\cite{Sedighi:03,DavidRus:09}.
 The authors of   Ref.~\cite{Sedighi:03} consider a two-state mean-field~\cite{Mean_field} description 
  that takes into account  cooperativity in binding of enzymes,
  and they discuss the bifurcation diagram, including the effects of  spatial dependence.
    In Ref.~\cite{DavidRus:09}, the results of a  stochastic simulation are  compared to those of a
    mean-field description that does  not explicitly consider spatial dependence.
   Perturbations due to cell divisions were considered, and instability of stable 
  steady states due to such perturbations were found in the stochastic simulation, but not the mean-field approach.
 It is an open question how to obtain 
 mean-field equations in the continuum starting from a stochastic description that predict the instabilities due to
  spatial dependance  that are
 observed in the microscopic simulations. Among other things, this is one of the questions that we address in this work.

 The considerable number of independently regulated modification sites
 in the chromatin  has been hypothesized to give rise to  a ``histone code"~\cite{Jenuwein:01}:
 There are  $2^T$ possible combinations of modified/unmodified configurations  of $T$ independently regulated
  PTMs, each of which potentially corresponds to a  distinct ``read-out" of information and ultimately a  different epigenetic 
 outcome.
Recent efforts in identifying  abundances of these histone  modification states 
 (also denoted as histone isoforms) have revealed that  only few of the 
  large number of possible isoforms are actually observed~\cite{Phanstiel:08,Pesavento:08}.
 It is also well known that regulation of different  PTMs is correlated. For example, 
  phosphorylation of H3 Ser10 stimulates acetylation of H3 Lys14 \cite{Lo:00}, and 
  methylation of  H3 Lys4 and Lys79 requires the ubiquitiniation of H2B Lys123 \cite{Sun:02,Ng:02}.
 In this work, we consider how such correlations in the regulation of PTMs reduce the information capacity of
  histone states. In particular, we study a model  that is motivated by an  interaction in the H3 N terminus where 
 Ser10 phosphorylation inhibits  Lys9 methylation~\cite{Rea:00}.

   We consider a master equation description of the  stochastic  dynamics of  histone  states (section~\ref{stochdyn}).
 The system consists of  a  large number of nucleosomes, 
   where each nucleosome  exhibits   several PTMs that are regulated by a particular class of enzymes. 
  We take into account 
   the reversible addition and removal of PTMs due to enzymatic  activity,
    as well as on-site (``local")  and  nearest-neighbour (``non-local")  feedback mechanisms  where modified nucleosomes recruit enzymes
     that   either act locally or  diffuse to adjacent nucleosomes.  
      We use a quantum many-body formulation 
       of the master equation \`{a} la Doi\cite{Doi:76}  and a tensor product state ansatz  to obtain a system of 
nonlinear difference equations (section~\ref{mapping}).
 We believe that the continuum limit of these  equations is a suitable
  mean-field description that captures the role of spatial dependance in the master equation.
 The reader who is not interested in the derivation of the nonlinear difference equations/partial differential equations can go directly
  to Eqs.~(\ref{PDE_system}), Eqs.~(\ref{ContPDE}) and Eqs.~(\ref{PDE_system2}).
 We numerically study the system of nonlinear partial differential equations (section~\ref{results}). 
    When considering one type of post-translational modification, and including at least two  modification sites, 
  bistable steady states are obtained without the necessity of explicit cooperativity at the level of the stochastic description
  (section~\ref{ODE}).
   The two stable steady states correspond to an unmodified state  and a state with a high number of PTMs.
 We observe  that increasing the number of  modification sites 
 increases the size of the  parameter regime where bistable steady states exist. 
 For a large number of
   modification sites, bistability is possible even if the 
coupling strength of  the  feedback mechanism is weak compared to the coupling strength of 
   local  processes. 
 We observe that the  spatial dependance due to the non-local feedback mechanism  leads to
    instabilities of steady states under certain spatial perturbations of the histone state (section~\ref{SpatialDependance}).
    These instabilities  manifest themselves in traveling wave solutions of 
 the system of nonlinear partial differential equations.
We also consider spatially dependent rate parameters, which arise from   adaptor 
 proteins, such as DNA binding transcription factors, that recruit histone modifying enzymes to specific regions 
 of chromatin (section~\ref{Hetparameters}). We discuss how such spatially dependent enzyme activity   gives rise to spatial heterogeneity in the epigenetic state. 
 Finally, we introduce a model of  two types PTMs that are regulated by different classes of enzymes and
 mutually inhibit each other (section~\ref{multiple_PTMs}). Such mechanisms are present in the chromatin, for example, in the case of  
 H3 Ser10 phosphorylation that inhibits  H3 Lys9 methylation~\cite{Rea:00}. 
   We find that inhibition in one direction is sufficient to reduce the full combinatorial set
    of four stable steady states to a set of three stable steady states where the presence of the  two types of PTM is mutually exclusive.
     We conclude by discussing open problems and future directions.


\section{Stochastic  dynamics of histone states}\label{stochdyn}

\begin{figure}[t]
\begin{center}
\includegraphics[width=7.cm,height=2.3cm]{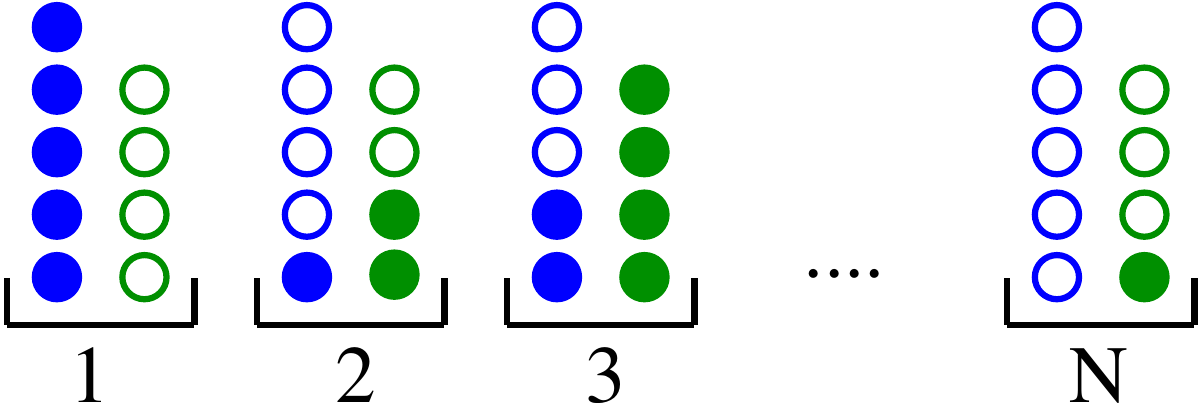}
\caption{Schematic illustration of an array of $N$ nucleosomes, each of which contains 
 $S^A=5$  PTMs A (blue) and $S^M=4$ PTMs  M (green) where PTMs of types A are regulated
  by a certain set of enzymes and PTMs  M are 
 regulated by another set of enzymes.
  Filled circles symbolize the presence of a PTM, empty circles indicate the absence of a PTM.
   In this example, occupations are $n_1^A=5$, $n_1^M=0$, $n_2^A=1$, $n_2^M=2$, etc.} 
\label{Nucleosomes}
\end{center}
\end{figure}

We consider a one-dimensional  array of $N$ nucleosomes.
Each nucleosome  contains several  modification sites 
of one or several independently regulated classes of PTMs,
  as schematically illustrated in Fig~\ref{Nucleosomes}.
    A system comprised of $N$ nucleosomes with  $S^A$  modification sites of type A (e.g., acetylation) 
         on each nucleosome 
  is described by a state
  $|n_1^A, n_2^A,..., n_N^A\rangle$ where  
  the number of modified (e.g., acetylated) sites on nucleosome $i$ is given by $n_i^A\in \{0,1,...,S^A\}$. 
   We denote by $P(n^A_1,n^A_2,...,n^A_N; t)$  the probability of finding the system in state
 $|n_1^A, n_2^A,..., n_N^A\rangle$   at time $t$.
  In this and the following sections, we shall restrict ourselves to a single class of PTMs (i.e., regulated by a particular
   set of enzymes); however, 
  in section~\ref{multiple_PTMs} we shall discuss the case of two types of PTM.

In the description of the stochastic dynamics of the histone state, we consider 
on-site (``local") and nearest-neighbour (``non-local" ) processes:
 \begin{enumerate}
 \item  The addition of a  PTM  $A$ at nucleosome $i$ with a rate $\lambda^A$,
 \begin{displaymath}
 n_i^A   \stackrel{\lambda^A}{\longrightarrow} n_i^A+1,
 \end{displaymath} 
 caused by enzymatic activity.
 \item The removal of a PTM  $A$ at nucleosome $i$  with a rate $\mu^A n_i^A$,
 \begin{displaymath}
 n_i^A \xrightarrow{\mu^A n_i^A}n_i^A-1,
 \end{displaymath}
 as a result of enzymatic activity.
 \item The addition of a  PTM  $A$  at nucleosome $i$  with a rate $  f(n^A_{i-1}, n_i^A, n^A_{i+1})$, 
  \begin{displaymath}
n^A_i\xrightarrow{f(n_{i-1}^A,n_i^A,n_{i+1}^A)}  n^A_i+1.
  \end{displaymath}
The choice 
\begin{equation}
f(n_{i-1}^A,n_i^A,n_{i+1}^A) =  \tilde{\alpha}^A  n_i^A + \alpha^A (n_{i-1}^A+n_{i+1}^A   -2 n_i^A) , 
\label{alpha_f}
\end{equation}
   corresponds to a  feedback mechanism  that is both local and non-local.
   The first term (coupling parameter $\tilde{\alpha}^A$) accounts for local feedback: 
   the more PTMs are present at nucleosome $i$,
   the more enzymes that add PTMs of type A (e.g., acetylases) are present at $i$, and the more likely is the addition of further PTMs of type A. 
   The second term (coupling parameter $\alpha^A$) corresponds to non-local feedback:
    the enzymes at nearest-neighbouring nucleosomes $i-1$ and $i+1$ diffuse to nucleosome $i$ and vice versa
     and, as in the case of local feedback, make the addition of additional PTMs more likely.
   
        \item The removal of a PTM $A$ at nucleosome $i$ with a rate $ n_i^A g(n^A_{i-1},n_i^A,n^A_{i+1})$, i.e.,
   \begin{displaymath}
   n^A_i\xrightarrow{n_i^A g(n_{i-1}^A,n_i^A,n_{i+1}^A)} n^A_i-1.
    \end{displaymath}
The choice 
\begin{equation}
g(n_{i-1}^A,n_i^A,n_{i+1}^A) =\tilde{\beta}^A (S^A- n_i^A) +   \beta^A (2n_i^A -n_{i-1}^A-n_{i+1}^A ),
\label{beta_g}
\end{equation}
corresponds to a  feedback mechanism  that is both local and non-local.
   The first term (coupling parameter $\tilde{\beta}^A$) accounts for  local feedback: 
   The fewer   PTMs are present at  nucleosome $i$ (i.e., the larger $S-n_i^A$),  the more enzymes that cause the removal of PTM $A$
  (e.g., deacetylases) are present at $i$, making the removal of further PTMs more likely. 
    The second term (coupling parameter $\beta^A$) corresponds to non-local feedback:
    the enzymes that cause the removal of PTMs A at nearest-neighbouring nucleosomes $i-1$ and $i+1$ diffuse to nucleosome $i$
    and vice versa
     and, as in the case of local feedback, make the removal of PTMs at site $i$ more likely.
 \end{enumerate}

         The master equation for the above  processes is given by
\begin{small}
 \begin{eqnarray}
&& \frac{dP(n_1^A,...,n_N^A;t)}{dt}=
\sum_{i=1}^N  [ \lambda^A  +f(n^A_{i-1},n_i^A,n^A_{i+1})   ][  P(n^A_1,...,n^A_{i-1},n_i^A-1,n^A_{i+1},...,n^A_N;t) -  
 P(n^A_1,...,n^A_{i-1},n^A_i,n^A_{i+1},...,n^A_N;t) ] \nonumber \\
 &&+ \sum_{i=1}^N [\mu^A+ g(n^A_{i-1},n_i^A,n^A_{i+1}) ][(n^A_i+1) P(n^A_1,...,n^A_{i-1},n^A_i+1,n^A_{i+1},...,n^A_N;t) -  n^A_i P(n^A_1,...,n^A_{i-1},n^A_i,n^A_{i+1},...,n^A_N;t) ].
\label{master_equation}
\end{eqnarray}
\end{small}
 

\section{Derivation of nonlinear difference equations}\label{mapping}

We shall now  introduce a  notation of the master equation (\ref{master_equation})
  that is motivated by quantum physics~\cite{Doi:76,Peliti:85}.
   Standard quantum physics notation is used, i.e.,
   $|n_1^A,..., n^A_N\rangle = |n_1^A\rangle\otimes |n_2^A\rangle \otimes ...\otimes |n^A_N\rangle$.
   We  define  
  \begin{displaymath}
  |\Psi(t)\rangle =\sum_{\{n\}} P(n^A_1,n^A_2,...,n^A_N;t) |n^A_1,n^A_2,...,n^A_N\rangle,
 \end{displaymath}  
    where the sum runs over all possible states. 
     We   introduce local raising and lowering operators~\cite{creation_annihilation}
  $\mathcal{R}_i$ and $\mathcal{L}_i$ that are defined by 
  \begin{displaymath}
   \mathcal{L}^A_i|n^A_i\rangle = n^A_i|n^A_i-1\rangle,   \hspace{1cm} \mathcal{R}^A_i|n^A_i\rangle = |n^A_i+1\rangle, \hspace{1cm}
  \mathcal{R}^A_i|S^A_i\rangle = 0,  \hspace{1cm} \mathcal{L}^A_i|0^A_i\rangle = 0, 
  \end{displaymath}
  Indices $A$  and $i$ of operators 
signify that the operators are applied to state $|n_i^A\rangle$. 
    When representing states $|0^A\rangle$, $|1^A\rangle$,..., $|S^A\rangle$   by the $S^A+1$ unit vectors in $S^A+1$ dimensions,
     the lowering and raising operators can be represented by $(S^A+1)\times (S^A+1)$ dimensional matrices,
     \begin{displaymath}
\mathcal{L}^A_i = \left ( \begin{array}{cccccc}  0&1&0&...&...&0\\ 0&0&2&0&...&0\\0&0&0&3&...&0\\...&...&...&...&...&...\\0&0&...&...&...&S^A\\
0&0&...&...&0&0
 \end{array}\right ),\hspace{2cm} \mathcal{R}^A_i=  \left ( \begin{array}{cccccc}  0&0&...&...&0&0\\ 1&0&...&...&...&0\\0&1&0&...&...&0\\...&...&...&...&...&...\\0&0&...&1&...&...\\
 0&0&...&...&1&0
 \end{array}\right )\ .
     \end{displaymath}
   The number operator is defined by $\mathcal{N}^A_i=\mathcal{R}^A_i\mathcal{L}^A_i = {\rm Diag}(0,1,2,...,S^A)$.
In this notation, the master equation becomes
\begin{equation}
\frac{\partial |\Psi(t)\rangle }{\partial t} = \mathcal{H} |\Psi(t)\rangle,
\label{QM_MasterEq}
\end{equation}
where $\mathcal{H}=  \mathcal{H}_1^A\otimes \mathcal{E}_2^A \otimes ...\otimes \mathcal{E}_N^A + \mathcal{E}_1^A \otimes \mathcal{H}_2^A\otimes ... \otimes \mathcal{E}_N^A +... +\mathcal{E}_1^A \otimes ...
\otimes \mathcal{H}_N^A $ (in simplified notation: $\mathcal{H} = \sum_{i=1}^N \mathcal{H}^A_i$), where
 $\mathcal{E}_i^A$ denotes the $S^A$-dimensional identity operator, and 
  \begin{eqnarray}
  \mathcal{H}^A_i &=& \lambda^A (\mathcal{R}^A_i  - \mathcal{I}^A_i) + \mu^A (\mathcal{L}^A_i - \mathcal{N}^A_i)  +
  ( \mathcal{R}^A_i  - \mathcal{I}^A_i)
 [ \alpha^A ( \mathcal{N}^A_{i-1} + \mathcal{N}^A_{i+1} -2\mathcal{N}^A_i  )+\tilde{\alpha}^A \mathcal{N}_i^A] 
 \nonumber \\
&&  + ( \mathcal{L}^A_i - \mathcal{N}^A_i)  
  [\beta^A  (\mathcal{M}^A_{i-1}  + \mathcal{M}^A_{i+1} -2\mathcal{M}^A_i)  + \tilde{\beta}^A \mathcal{M}_i^A],
   \label{QM_H}
   \end{eqnarray}
where $\mathcal{I}^A ={\rm Diag}(1,1,...,1,0)$, and
$\mathcal{M}^A={\rm Diag}(S^A,S^A-1,...,1,0)$.
 In (\ref{QM_H}), we substituted  the functions (\ref{alpha_f}) and (\ref{beta_g}). 
We note that (\ref{QM_MasterEq}) is   an imaginary-time Schr\"odinger equation. The 
 system corresponds to a quantum spin chain, though with a non-hermitian Hamitonian. 

The master equation (\ref{QM_MasterEq}) is equivalent to a functional variation~\cite{Eyink:96},
\begin{equation}
    \frac{\delta  \Gamma}{\delta \Phi} =0,
            \label{functional_variation}
    \end{equation}
    where 
    \begin{displaymath}
 \Gamma = \int dt \langle \Phi|(\partial_t -\mathcal{H})|\Psi\rangle.
  \end{displaymath}
 Since the system can be viewed as a quantum spin chain,  albeit with a  non-hermitian Hamiltonian $\mathcal{H}$, 
      we make 
   an     ansatz for the wave-function in the Schr\"odinger picture as a tensor product state, 
        \begin{equation}
       | \Psi(t)\rangle=\prod_{i=1}^N |\Psi_i(t)\rangle, 
          \hspace{2cm}\langle \Phi|=\prod_{i=1}^N \langle \Phi_i|\ .
                 \label{ansatz1}
        \end{equation}
        and we write $|\Psi_i(t)\rangle$ as a superposition of all possible states (we shall drop indices $A$ from this point on),
     \begin{equation}
|\Psi_i(t)\rangle 
 = \sum_{n=0}^S C_{i,n}(t) |n\rangle=  \left (\begin{array}{c} C_{i,0}(t) \\C_{i,1}(t)\\.\\.\\.\\C_{i,S}(t)\end{array}\right ),
\hspace{1cm}  \langle \Phi_i| = \sum_{n=0}^S \langle n| e^{\phi_{i,n}} =   (e^{\phi_{i,0}}\;\;e^{\phi_{i,1}}\;\;...\;\;e^{\phi_{i,S}}),
\label{ansatz2}
        \end{equation}
        where $\sum_{n=0}^S C_{i,n}=1$, and $C_{i,n}$ denotes the probability that nucleosome $i$ has $n$ modified sites.
         Since $\sum_{n=0}^{S} C_{i,n}=1$, 
this ansatz obeys the probabilistic constraint $\langle \Phi|\Psi\rangle|_{\phi_{i,n}=0}=1$ (i.e., expectation values
         $\langle \Phi|O|\Psi\rangle$ of an observable $O$ are properly normalized).

       Using this ansatz, the master equation in the formulation of (\ref{functional_variation})
becomes 
    \begin{equation}
   \left (  \left \langle \frac{\partial \Phi}{\partial \phi_{i,k}} \right | \left .  \frac{\partial \Psi}{\partial C_{i,n}} \right \rangle\frac{dC_{i,n}}{dt}
       -    \left \langle \frac{\partial \Phi}{\partial \phi_{i,k}} \left  | \mathcal{H} \right  |  \Psi \right \rangle \right )_{\phi_{i,k=0} }= 0.
       \label{variational}
     \end{equation}
    Evaluating (\ref{variational}) yields a system of nonlinear difference equations
for the probabilities $C_{i,n}$ that the nucleosome  $i$ has $n$ modifications,
       \begin{eqnarray}
\frac{dC_{i,0}}{dt} &=& -\lambda C_{i,0}+\mu C_{i,1} - 
C_{i,0} (
 \alpha  F^{\nabla}_i + \tilde{\alpha} \langle n_i\rangle )
+ C_{i,1} ( \beta G^{\nabla}_i +\tilde{\beta}  \langle m_i\rangle ),
  \nonumber  \\
\frac{dC_{i,n}}{dt}&\stackrel{1\le n  < S}{=}& -\lambda( C_{i,n}-  C_{i,n-1}) -\mu (n C_{i,n} -  (n+1) C_{i, n+1})\
- ( C_{i,n} -C_{i,n-1} ) (\alpha F^{\nabla}_i + \tilde{\alpha } \langle n_i\rangle)  \nonumber \\
 &&- ( n C_{i,n}- (n+1) C_{i,n+1}) (\beta G^{\nabla}_i  + \tilde{\beta} \langle m_i\rangle ) ,
\nonumber \\
\frac{dC_{i,S}}{dt} &=& \lambda C_{i, S-1} - S \mu C_{i,S} 
+C_{i,S-1}  (\alpha F^{\nabla}_i +\tilde{\alpha} \langle n_i\rangle )
- S  C_{i,S} (\beta G^{\nabla}_i+\tilde{\beta} \langle m_i\rangle), 
\label{PDE_system}
\end{eqnarray}
where 
\begin{displaymath}
\begin{array}{llllll}
F^{\nabla}_i = \langle n_{i-1}\rangle-2\langle n_i \rangle + \langle n_{i+1}\rangle &{\rm if } \;\;\;1< i < N,  \hspace{1cm} &  F^{\nabla}_1 =-2\langle n_1\rangle +  \langle n_{2}\rangle  \hspace{1cm} &
F^{\nabla}_N =  \langle n_{N-1}\rangle -2\langle n_N \rangle ,
\\[2mm]
G^{\nabla}_i=\langle m_{i-1}\rangle-2\langle m_i\rangle + \langle m_{i+1}\rangle & {\rm if} \;\;\; 1< i < N, \hspace{1cm} &  G^{\nabla}_1 =  -2\langle m_1\rangle +\langle m_{2}\rangle
\hspace{1cm} &
G^{\nabla}_N =  \langle m_{N-1}\rangle -2\langle m_N\rangle , &
\end{array}
\end{displaymath}
and
\begin{eqnarray}
\langle n_{i}\rangle &=& \sum_{n=0}^S  n C_{i,n}\label{exp_n}\\
\langle m_{i}\rangle& =& \sum_{n=0}^S (S-n) C_{i,n}=S-\langle n_i\rangle\label{exp_m},
\end{eqnarray}
(open boundary conditions).

  Equations (\ref{PDE_system}) are a discretization of a system of nonlinear
  reaction-diffusion equations. Let
  $\ell_0$ be the lattice spacing (distance between nucleosomes). In the mean-field/continuum limit $\alpha\to \alpha/\ell_0^2$, 
  $\beta\to\beta/\ell_0^2$, and $\ell_0\to 0$, we obtain the system of
 nonlinear partial differential equations for variables $C_n(x,t)$, $n=0,1,...,S$,
 \begin{eqnarray}
 \frac{\partial C_0}{\partial t} &=& -\lambda C_0 +\mu C_1 -  C_0\left ( \alpha \sum_{s=0}^S \left  (  s \frac{\partial^2C_s}{\partial x^2} \right )+\tilde{\alpha} \sum_{s=0}^S (s C_s)\right )
  +C_1 \left ( \beta \sum_{s=0}^S \left ( (S-s) \frac{\partial^2C_s}{\partial x^2} \right ) +\tilde{\beta}   \sum_{s=0}^S( S-s C_s)\right ) ,
   \nonumber \\[1mm]
 \frac{\partial C_n}{\partial t} &\stackrel{1\le n < S}{=} & -\lambda (C_n-C_{n-1} ) -\mu(nC_n-(n+1)C_{n+1}) - (C_n-C_{n-1}) \left (
\alpha \sum_{s=0}^S \left (  s \frac{\partial^2C_s}{\partial x^2}  \right ) + \tilde{\alpha} \sum_{s=0}^S (s C_s ) \right )
   \nonumber \\
   && - (nC_n - (n+1) C_{n+1}) \left  ( \beta \sum_{s=0}^S \left ( (S-s) \frac{\partial^2C_s}{\partial x^2}  \right )   + \tilde{\beta} \sum_{s=0}^S (S-s C_s )\right )
   \label{ContPDE} \\[1mm]
 \frac{\partial C_S}{\partial t} & =&\lambda C_{S-1}  -S\mu C_S + C_{S-1}\left ( \alpha  \sum_{s=0}^S   \left (  s \frac{\partial^2C_s}{\partial x^2}   \right )
 +\tilde{\alpha}    \sum_{s=0}^S  (sC_s )\right ) -   S C_S \left (  \beta  \sum_{s=0}^S  \left ((S-s) \frac{\partial^2C_s}{\partial x^2} \right ) +\tilde{\beta} 
  \sum_{s=0}^S (S-s C_s )\right ).
\nonumber 
 \end{eqnarray}
The diffusion terms are multiplied with the probabilities themselves.
 We note that the coefficient in front of the diffusion term is degenerate, and 
 it is of interest to 
rigorously show the existence and stability of traveling wave solutions in reaction-diffusion equations of this type.

\section{Results}\label{results}

In what follows, our analysis is based on numerical analysis of the system (\ref{PDE_system}) over
a finite parameter range. 
In the following, we set parameters $\tilde{\alpha}=4\alpha$ and $\tilde{\beta}=4\beta$, and 
 we emphasize that  varying  the relative strength of local and non-local feedback does not qualitatively affect the results of our study. 
We note that as long as one is interested in the asymptotics (asymptotically long
 time) behavior of solutions of difference equations, what matters as input in the equations is the ratio 
 (relative strength) of various coupling parameters (e.g., $\beta/\alpha$, $\lambda/\alpha$, etc.). 
  One can always divide by a non-zero coupling parameters and rescale time to absorb this parameter in the 
 left-hand-side of the difference equations.

In section IV A, we will first discuss bistability in the model while neglecting spatial dependence.
  We will then incorporate spatial effects in part B, which we note fundamentally alters the picture.
  In section~\ref{Hetparameters}, we discuss the effects of spatial heterogeneity and in section~\ref{multiple_PTMs},
   we discuss multiple correlated PTMs.

\subsection{Multiple stable steady states in the $S$-state model and the role of $S$ }\label{ODE}

In this section, we discuss the results of the  nonlinear difference equations  (\ref{PDE_system}) 
 when neglecting the 
 spatial dependance,  
 i.e., $C_{1,n}=C_{2,n}=...=C_{N,n}=C_n$. 
 In this case, a system of coupled nonlinear
 ordinary differential equations (ODE) is obtained,
 
 \begin{eqnarray}
 \frac{dC_0}{dt} &=&-\lambda C_0 +\mu C_1 - 4\alpha C_0^2 +4\beta C_0 C_1,  \nonumber \\
 \frac{dC_n}{dt} &\stackrel{1\le n<S}{ =} & -\lambda (C_n-C_{n-1}) - \mu (nC_n-(n+1)C_{n+1})- 4\alpha C_n ( C_n  - C_{n-1} )  - 4\beta C_n  (nC_n-(n+1)C_{n+1}), \nonumber \\
 \frac{dC_S}{dt} &=& \lambda C_{S-1} -S\mu C_S +4\alpha C_{S-1}C_S -4S\beta C_S^2.
 \label{ODE_system}
 \end{eqnarray}
Using this simplified ODE description, we
 evaluate steady states    by setting $dC_{n}/dt=0$,  and study their stability by analyzing the Jacobian matrix.
Expressions for the steady state probabilities $C_n$  as a function of parameters
 $\lambda$, $\mu$, $\alpha$ and $\beta$ can be evaluated analytically. However, the resulting expressions are
cumbersome and increasingly difficult to obtain for increasing $S$, and therefore 
 calculations have been done  numerically over a finite parameter range. 
 
    \begin{figure}[t]
\begin{minipage}[t]{0.47\textwidth}
\begin{center}
\includegraphics[width=9.4cm]{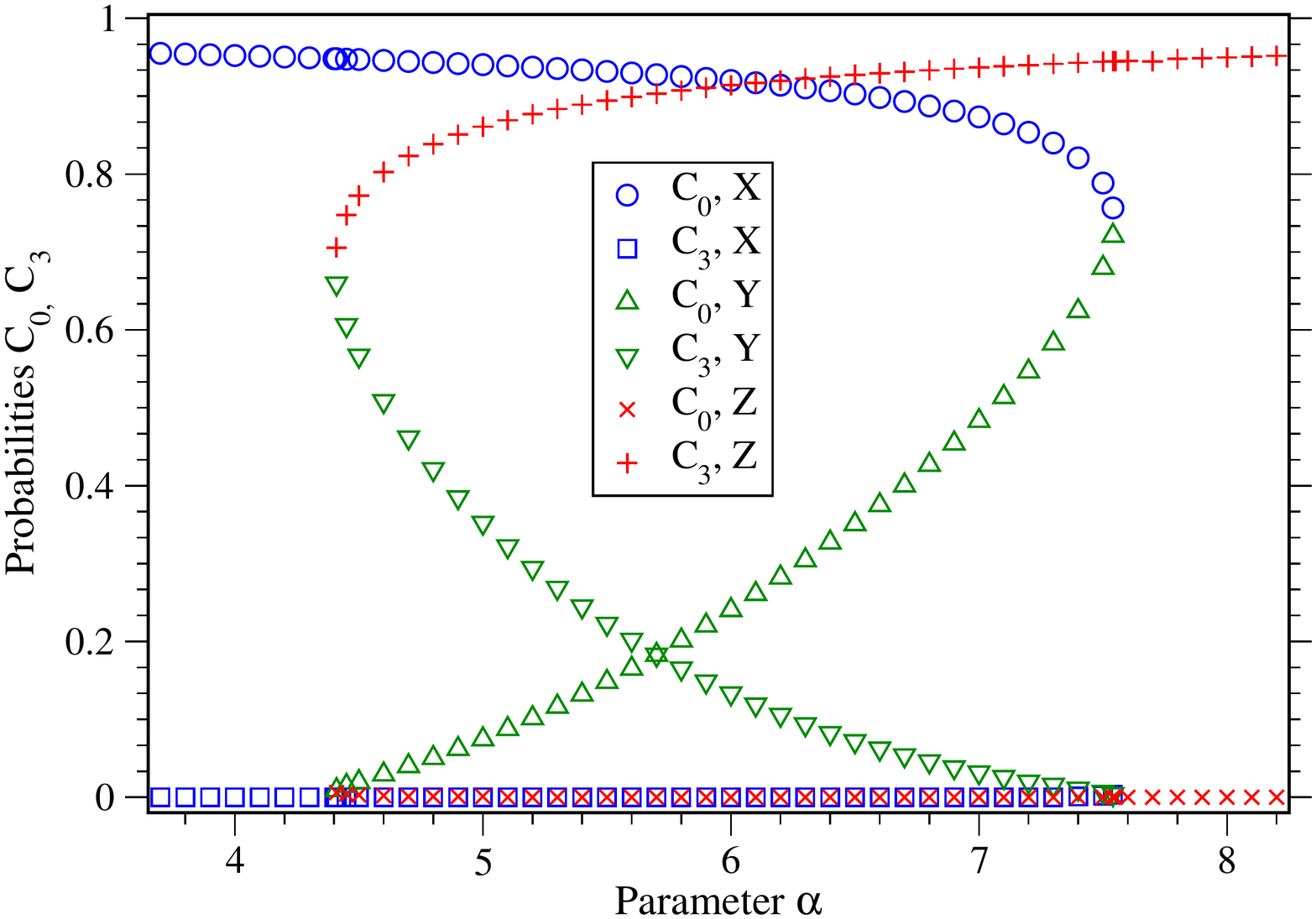} 
\caption{
Bifurcation diagram showing the steady state probabilities $C_0$ and $C_S$ for  $S=3$  modification sites  
 and parameters  $\lambda=\mu=1$, $\beta=3$
as  a function of parameter $\alpha$. 
We denote  the stable steady states where $C_0\approx 1$ (few PTMs) and $C_S \approx 1$ 
(large number
 of PTMs) by X and  Z,
 respectively, and  the unstable steady state by Y.
For  $\alpha\in [4.4,7.5]$,  steady states X, Y and Z appear, while for small $\alpha$ only
 X persists  and for large $\alpha$ only Z persists. 
  }
 \label{4State_lin_bifur}
\end{center}
\end{minipage}\hfill
\begin{minipage}[t]{0.47\textwidth}
\begin{center}
\includegraphics[width=9.4cm]{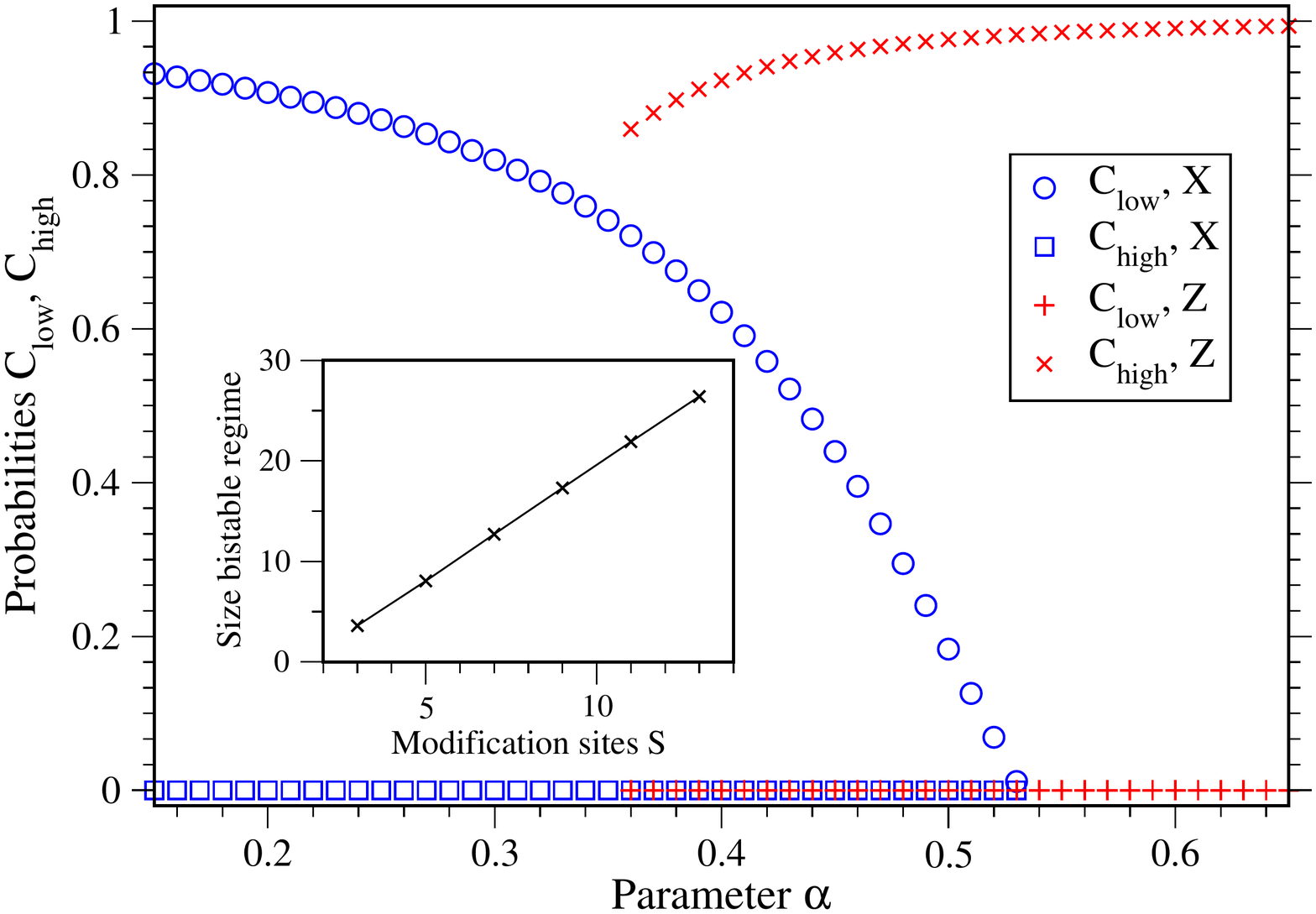} 
 \caption{Bifurcation diagram for $S=50$ modification sites  showing the steady state probabilities $C_{\rm low} =\sum_{n=0}^4 C_n$ (i.e., 
 low number of PTMs) and $C_{\rm high} =\sum_{n=46}^{50} C_n$ (i.e., high number of PTMs) of the stable steady states
  X and Z
 as a function of $\alpha$. The remaining parameters
  are $\lambda=5$, $\mu=1$, $\beta=0.01$.
   For $\alpha\in [0.36,0.53]$, bistability persists.
   Note that $\alpha\ll \lambda$ and $\beta \ll \mu$.
Inset:  Width of the bistable regime in units of  $\alpha$ as a function of
the number of modification sites  $S$ ($\lambda=\mu=1$, $\beta=3$).
It can be seen to increase linearly. 
 }
  \label{multistable_range}
\end{center}
\end{minipage}\hfill
\end{figure}

For more than one modification site, i.e., $S\ge 2$,  and appropriately chosen  parameters (see below) 
 we find that a  parameter regime exists where  three steady states coexist.
 The multistability is a consequence of the nonlinearities in Eqs.(\ref{ODE_system}) 
 that are introduced by the feedback 
  terms.
 Two of the steady states are  stable 
   attractors  and one steady state is an unstable saddle point. 
   We  note that no explicit cooperativity is required in order to obtain bistability if $S$ is chosen larger or equal than 
    two.
   
The bistability is illustrated  in the  bifurcation diagram of Fig.~\ref{4State_lin_bifur} where the steady state probabilities
       $C_0$ and $C_3$ are shown as a function of parameter $\alpha$ (the parameters used  are $S=3$,  $\mu=\lambda=1$, $\beta=3$).
   If the  feedback term for  enzymes that catalyse the addition of PTMs is weak compared to 
       the  feedback term  of enzymes that catalyse the removal of PTMs,
        only one steady state appears, as can be seen in Fig.~\ref{4State_lin_bifur} for  $\alpha<4.4$. 
        This steady state, which we denote by X, is  characterized by $C_0\approx 1$, i.e., 
        it corresponds to a state where very few PTMs are present. 
      If the effects of the  two 
   terms that add PTMs approximately  are roughly equal  to the effects of the  two terms that remove PTMs, three steady states exist
    ($\alpha\in [4.4, 7.5]$ in Fig.~\ref{4State_lin_bifur}). 
    In addition to steady state X, a steady state with $C_S\approx 1$  appears. This steady state corresponds 
    to a state with a high number of PTMs, and we shall denote it by Z.
  A third steady state  (denoted by Y in Fig.~\ref{4State_lin_bifur}) is unstable.
  Finally, for large enough $\alpha$, only steady state Z persists, as illustrated in Fig.~\ref{4State_lin_bifur} for $\alpha>7.5$.

We note that in the previous paragraph we referred
 to the ``strengths" of the four terms (1.-4. in section~\ref{stochdyn})
as they can be read from the expectation values, e.g., 
       $\langle \Phi | \sum_i f(n_{i-1},n_i,n_{i+1})|\Psi\rangle$.
   In contrast, in the following paragraph, we shall  refer to the magnitudes of the coupling parameters 
   (i.e.,  $\alpha$, $\beta$, $\mu$, $\lambda$) themselves.
   The  values of the coupling parameters are  controlled externally (e.g., the concentration, catalytic rate and diffusion rate of enzymes), while 
   the expectation values also depend on system-dependent parameters (i.e., the number of modification sites $S$).

Bistability is obtained only if   
  both   feedback terms are present, i.e., if both $\alpha$ and $\beta$ are non-zero.
   If the number of modification sites, $S$,  is small, bistable steady states appear only if 
     the coupling parameters of the  feedback terms are large  compared to those of  the local terms, i.e., 
      only if the ratios $\lambda/\alpha$ and $\mu/\beta$ are small enough. 
 However,  with increasing number of modification sites $S$, the size of the parameter regime 
       where multiple steady states appear increases, as shown in the inset of   Fig.~\ref{multistable_range}, and for
       large enough $S$, bistability can be established even if $\alpha \ll \lambda $ and $\beta \ll \mu$, as shown in Fig.~\ref{multistable_range}.
     The existence of a large  number of modification sites $S$
    that are regulated by a particular set of enzymes      thus 
  allows for a larger parameter regime of bistability.



\subsection{Spatial dependance}\label{SpatialDependance}

   \begin{figure}[t]
    \begin{minipage}[t]{0.47\textwidth}
     \begin{center}
           \includegraphics[width=9.4cm]{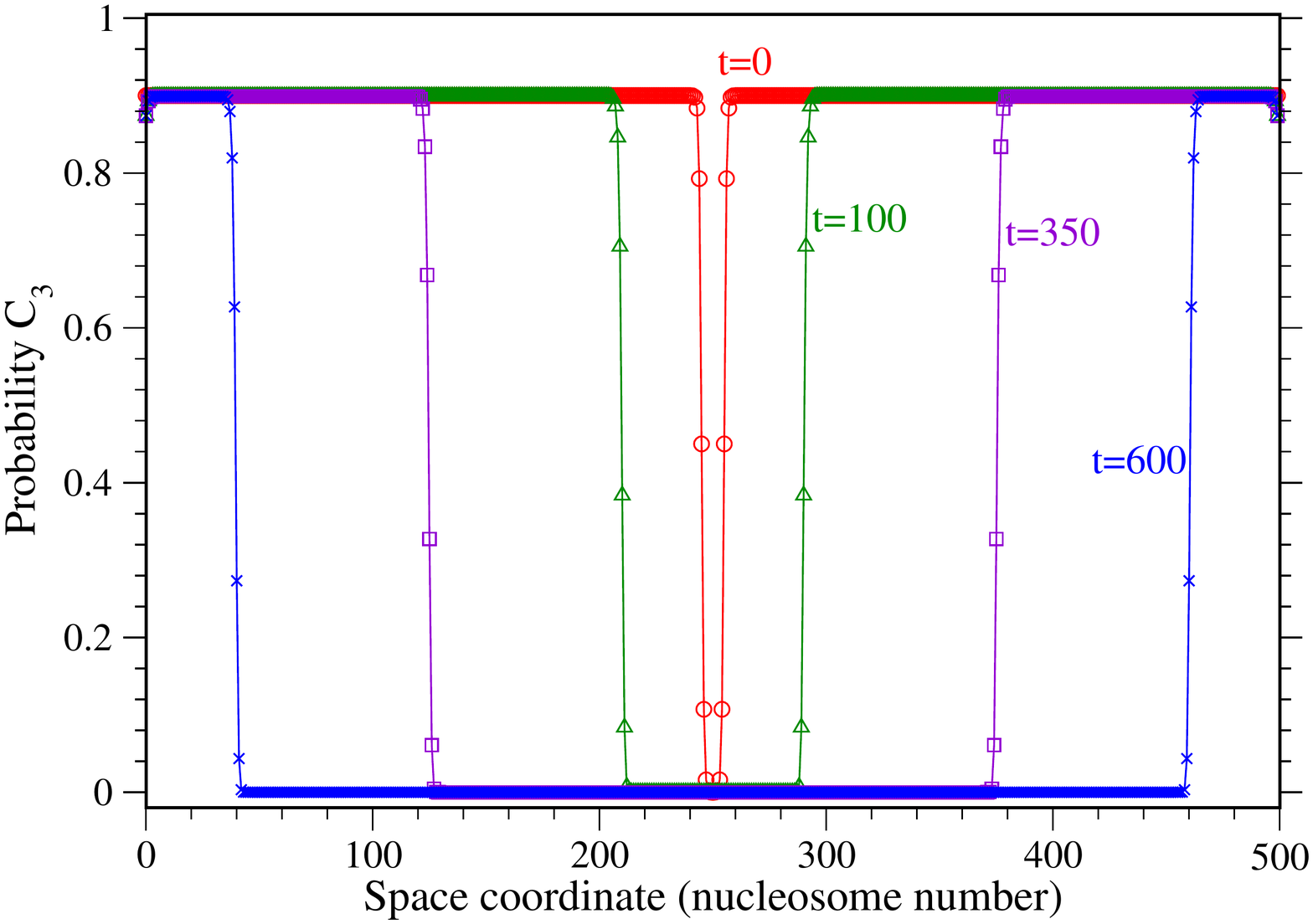}
   \caption{ Time evolution of  probabilities $C_{i,3}(t)$.
The   system ($S=3$ modification sites) is initially (time $t=0$, red curve) in steady state Z (where $C_3\approx 0.9$), except for
  few nucleosomes in the center that are strongly perturbed and whose probabilities are in the domain of steady state X.
  Parameters are $\lambda=\mu=1$, $\beta=3$, $\alpha=5.6$.
Two traveling wave fronts move towards the boundaries and drive the system into steady state X.
  The velocity of the waves is constant.}
  \label{TW_instable1}
   \end{center}
\end{minipage}\hfill
   \begin{minipage}[t]{0.47\textwidth}
   \begin{center}
\includegraphics[width=9.4cm]{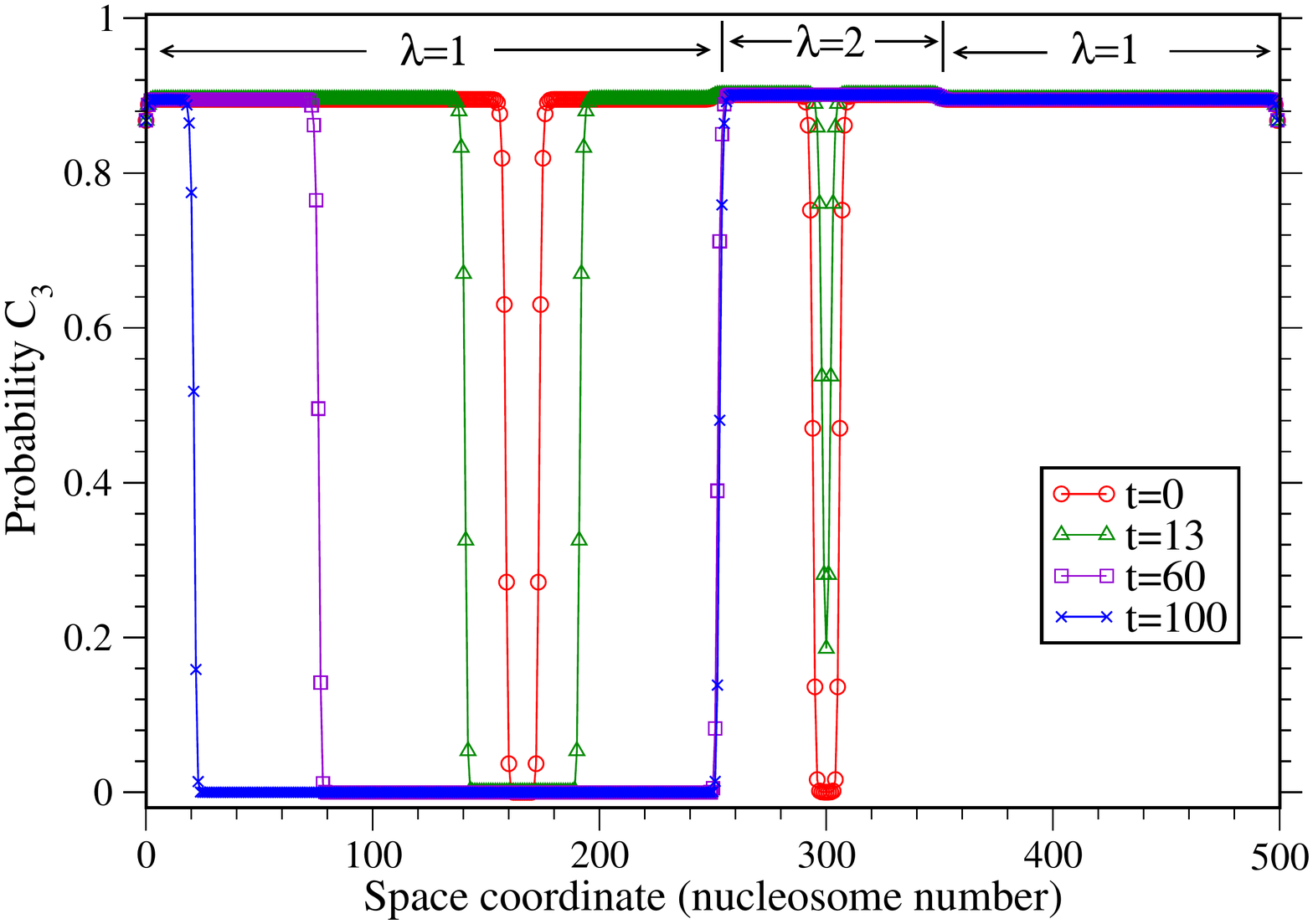}
\caption{
 Time evolution of   probabilities $C_{i,3}(t)$. Parameters are as in Fig.~\ref{TW_instable1},
  except that  $\lambda=1$ in part of the system, and  $\lambda=2$ 
in the remainder, as indicated. The   system is initially 
 in steady state Z 
 except for
 few nucleosomes in both regions whose states lie in the domain of X (red circles). 
In the left region ($\lambda=1$), 
   two wave fronts move towards the boundaries, however, once the right front hits the $\lambda=2$ region,
   it is stopped. In the $\lambda=2$ region, the perturbation does not cause the system to approach X.  The reason is that 
  for $\lambda=1$, steady state X is the ``stronger attractor", while for
    $\lambda=2$, Z is the ``stronger attractor" (terminology see text). }
  \label{TW_instable3}
\end{center}
    \end{minipage}\hfill
\end{figure}

In this section we will explicitly take into account spatial dependence, which is incorporated in the solutions to equations (\ref{PDE_system}).
We numerically integrate  (\ref{PDE_system}) and find that
 the stable steady states that were discussed in the previous section  
 may become unstable for certain initial conditions. 
 We illustrate this in  Fig.~\ref{TW_instable1}: We  set the initial probabilities 
  $C_{i,n}$  of  the nucleosomes to those of steady state $Z$ (the steady state where $C_{i,S}$ is large),
   except for very few nucleosomes  where we set the initial probabilities to  values close to those corresponding to the second steady state X
   \cite{Initial_state}.
   It can be seen that the system approaches steady state $X$, i.e., the spatially  restricted perturbation of the 
   histone state  causes instability.
 This instability manifests itself by  traveling wave solutions of the  system of equations (\ref{PDE_system}). 
It can be seen in Fig.~\ref{TW_instable1}  that for a perturbation away from the boundaries,
 two traveling wave fronts develop which travel at a constant velocity
 towards the boundaries of the system.
If the perturbation is located at one of the boundaries   of the system,  only one wave front  develops.

 There exists a set of parameters  $S$, $\lambda$, $\mu$, $\alpha$ and $\beta$ where the velocity of the traveling wave(s) is zero.
  At that point,  both steady states, X and Z,  are stable with respect to spatial perturbations.
  For the parameters set of Fig.~\ref{TW_instable1}, this transition occurs at
  $\alpha^*\approx 5.7$ (bistability occurs for $\alpha\in[4.4,7.5]$).
For $\alpha<\alpha^*$ and within range of bistability, the steady state X is the ``stronger attractor":
 If the initial state is Z and  
 at least one nucleosome is  perturbed such that 
  its state  is in the domain of  fixed point   X, the system approaches X, as is illustrated in Fig.~\ref{TW_instable1}.
 If the initial state is X, and  at least one nucleosome is  perturbed such that  its state
 in the domain of steady state Z, the system
   bounces back into steady state X. 
   In  contrast, for $\alpha>\alpha^*$, steady state Z is the ``stronger attractor":
   If the initial state is X and at least one nucleosome is perturbed such that  its state is in the domain of Z, the system approaches Z.
If the initial state is Z, and at least one nucleosome is  perturbed such that its state is in the domain of 
attraction of  X, the system
   bounces back into steady state Z.
   
     In conclusion,  for parameters  $\alpha<\alpha^*$,   steady state X exhibits a very high degree of stability 
as any initial state of the system that gives rise to  traveling wave solutions yields traveling waves that drive the system into
 state X. 
 In contrast,  for parameters  $\alpha>\alpha^*$, any traveling wave solution will drive the system into steady state Z.
We note that when the asymptotic behaviour of equations (\ref{PDE_system}) are considered, the number of nucleosomes 
in the system is not relevant.  However, a larger number of nucleosomes does result in a longer duration 
for the traveling wave to spread over the entire system, which may be relevant if intermediate time scales are considered.

 Instabilitities due to traveling wave solutions could have significant impact on the stability and inheritance of 
chromatin steady states in daughter cells upon division. During cell division, it is thought that the parental nucleosomes 
are randomly distributed among the two daughter cells, with the second half being newly synthesized~\cite{Annunziato:05}.
   The modification state of these new nucleosomes is crucial to the stability of the epigenetic state in the presence of
  non-local feedback terms. This can be seen as follows. The  cell division can be modeled 
  by  replacing  the states of half of the nucleosomes
  (randomly selected) at periodic intervals. Assume that the system is initially in steady state Z and parameters are
  set to the values of Fig.~\ref{TW_instable1} where X is the ``stronger attractor".
  If the states of the newly synthesized nucleosomes are random (i.e., any state is possible), some of these nucleosomes
  might be in states that are in the domain of steady state X right after cell division.
In this case, a traveling wave can form, and drive the system into steady state X (after 
    one, several or many divisions, depending on the time-scales involved).
   We have verified this numerically.
 However, if the states of the newly synthesized nucleosomes are  correlated with the state of the nucleosomes in the
 mother cell such that the states of the new nucleosomes are in the  domain of attraction of the original state, such instabilities
  cannot arise. In the presence of non-local effects, a sufficient correlation between mother and daughter nucleosome states 
  is hence necessary to  preserve the chromatin state.
This would relate to the notion of epigenetic memory and in fact there is a relation between
    daugher cell state and mother state [an example was discussed in the second paragraph of the introduction]. 
    However, how this is conveyed at the molecular level remains a challenging open question.

We   conclude this section with a short discussion of the effects of considering explicit cooperative behaviour
 in the feedback terms. 
Explicit  cooperative action of enzymes on-site, as well as of enzymes on nearest-neighbouring nucleosomes 
   can be implemented using ansatz 
   $f^{\rm coop}(n_{i-1},n_i,n_{i+1})  = f(n_{i-1},n_i,n_{i+1})+\delta  n_{i-1}n_i n_{i+1} $
   and 
   $g^{\rm coop}(n_{i-1},n_i,n_{i+1})  = g(n_{i-1},n_i,n_{i+1})+ \gamma (S-n_{i-1})(S-n_i)(S-n_{i+1})$
      Using the approach of sections~\ref{stochdyn},~\ref{mapping} and ~\ref{ODE}, bistable steady states are observed,
   as was the case for the model without explicit cooperative action. However, bistability is possible even for the case $S=1$.
           This in agreement with prior
    studies of two-state models with explicit cooperativity~\cite{Sedighi:03,DavidRus:09}.
   The difference equations that are obtained using this ansatz, or their continuum version, 
 admit traveling wave solutions, as in the case of our model without explicit cooperative behaviour (\ref{PDE_system}) where $S\ge 2$.


 \subsection{Spatially heterogeneous enzymatic activity }\label{Hetparameters}

 In biological systems, nucleosome modifying enzymes are typically recruited to specific
  regions of the chromatin by adaptor proteins, such as DNA-binding transcription factors.  As a result, 
  the activity of these enzymes depends on the region of the chromatin.
 The increased or decreased  activity  of enzymes at certain nucleosomes can be taken into account
 by including a spatial dependance in  parameters $\lambda$ and $\mu$, i.e., $\lambda_i$ and $\mu_i$, 
 where $i$ is the nucleosome number.
  At each space point, the steady states are determined by the respective $\lambda_i$ and $\mu_i$, 
  i.e., the steady states locally correspond to the steady states with homogenous activity.
   Hence the parameter regimes where multiple stable steady states appear vary in size and position,
    and steady state probabilities $C_{i,n}$ also depend on the nucleosome number $i$.
    For example, when choosing $S=3$, $\mu=1$, $\beta=3$, and $\lambda=1$, bistability exists for $\alpha\in[4.4,7.5]$ 
    and $\alpha^*\approx 5.7$, while for 
parameters  $S=3$, $\mu=1$, $\beta=3$ and      $\lambda=2$, bistability persists for $\alpha\in[4.3, 6.9]$, where $\alpha^*\approx 5.5$.
 As a consequence, for parameter $\alpha=5.6$, steady state X is the stronger attractor (in the sense explained in 
 section~\ref{SpatialDependance})   for the former choice of parameters, 
 while steady state Z  is the stronger attractor for the latter choice of parameters. 
   When perturbing a system that is initially in steady state Z in both $\lambda$-regions, 
 traveling wave solutions drive the system into steady state X at the nucleosomes where 
  $\lambda=1$, but not in regions where $\lambda=2$, and the traveling waves
   in the region where $\lambda=1$  are  stopped once they hit regions where $\lambda=2$, as shown in Fig.~\ref{TW_instable3}.
  Spatial dependence on the activity of histone modifying enzymes that is conferred by 
 recruitment to regulatory regions of chromatin by transcription factors may thus stabilize 
 the histone state from local and non-local perturbation.


   \begin{figure}
   \begin{center}
        \includegraphics[width=10.5cm]{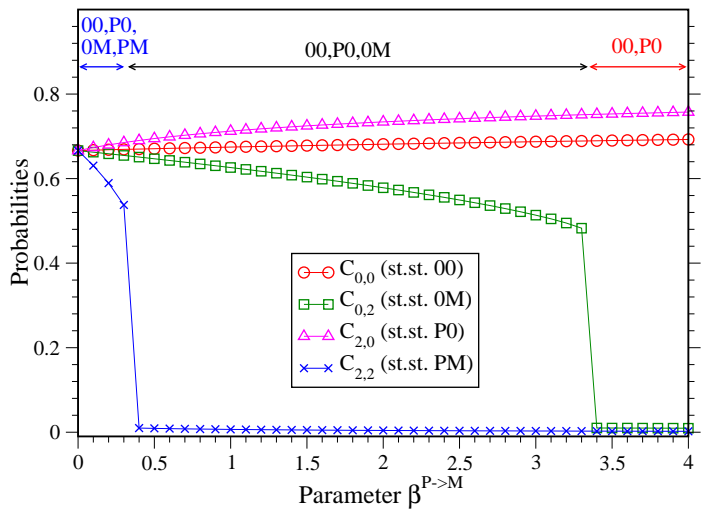}
   \caption{
   Bifurcation diagram showing  steady state probabilities $C_{p,m}$ 
 of stable steady states for model (\ref{PDE_system2}) with two types of modifications, labeled by P and M, 
    as a function of parameter $\beta^{P\to M}$. Note that only P inhibits M, but not vice versa, i.e., $\beta^{M\to P}=0$. The remaining parameters are given by
     $S^P=S^M=2$, $\lambda^P=\mu^P=\lambda^M=\mu^M=1$, $\beta^P=\beta^M=3$, $\alpha^P=\alpha^M=4.5$.
  For very small $\beta^{P\to M}$, four stable steady states (st.st.) exist, labeled by 
  00 (low P and low M), P0 (high P, low M), 0M (low P, high M), and PM (high P and high M).
In an intermediate parameter regime, $\beta^{P\to M}\in [0.4,3.3]$, stable steady states 00, P0 and 0M persist,
while for $\beta^{P\to M}>3.3$, only steady states 00 and P0 appear.
We note that inhibition in only {\it one} direction (as $\beta^{M\to P}=0$) is sufficient to 
 obtain a parameter regime where steady states have either a high number of PTMs P or M, or neither, but not both.}
  \label{Two_type_bifur}
   \end{center}
\end{figure}

\subsection{Multistability in model of several types of correlated  PTMs}\label{multiple_PTMs}

 Most proteins, such as histones, that are subject to PTM-dependent regulation are regulated via multiple
modifications.
In this context, we discuss the results of  including  several  types of modifications where each type is associated with 
 different sets of enzymes, and thus different rate parameters $\lambda$, $\mu$, $\alpha$, $\beta$, $\tilde{\alpha}$ and $\tilde{\beta}$.
  For example, one might consider different classes of acetylation (or phosphorylation, ubiquitination, etc.) sites, each of them associated with a  different enzyme. 
Alternatively, one might consider  PTMs of type P (e.g., phosphorylation) and PTMs of type M (e.g., methylation),
 with different rate parameters, $\lambda^P$ and $\lambda^M$, $\alpha^P$ and $\alpha^M$, etc.
We denote by $C_{i,p,m}$ the probability of finding nucleosome $i$ in the state with $p$ PTMs of type P and
 $m$ PTMs of type M. In this model, the number of stable steady states is four:
   the number of both M and P modifications is high (labeled by PM in the following),
    the number of P modifications is high and the number of M modifications is low (labeled by P0),
    the number of M modifications is high and the number of P modifications is low (labeled by 0M),
and  the number of both M and P modifications is low (labeled by 00).
 More generally, for $T$
 independent classes of modification sites, where a particular class of sites is associated with a particular set of coupling parameters,
 $2^T$ stable steady states are obtained. These steady states  correspond to  all possible combinations of  states of high and low 
  numbers of PTMs, i.e., all possible binary strings of
  length $T$.
  
 In practice, however, different types and sites of PTMs are often not independent from each other.
There are examples where the presence of a certain PTM inhibits the addition of another PTM.
 An example is the H3 N teminus where Ser10 phosphorylation inhibits
 Lys9 methylation~\cite{Rea:00}. 
In the following, we derive difference equations using the formalism introduced in sections~\ref{stochdyn} and \ref{mapping} 
for a model of
 two types of PTMs, P and M,  that mutually inhibit each other. 
We consider the processes 1.-4. (section~\ref{stochdyn}) 
separately for each of the two PTMs 
 and add mutual inhibition (note that  $m\equiv n^M$, $p\equiv n^P$):
\begin{eqnarray}
n_i^P n_i^M  \stackrel{ \beta^{P\to M} n_i^Pn_i^M }{\longrightarrow} n_i^P (n_i^M-1 ),\label{pm} \\
n_i^P n_i^M  \stackrel{ \beta^{M\to P} n_i^P n_i^M}{\longrightarrow} (n_i^P-1 )n_i^M.\label{mp}
\end{eqnarray} 
In the case of (\ref{pm}), the presence of PTMs of type P leads to the removal of   PTMs of type M, and in the 
case of (\ref{mp}), the presence of PTMs of type M leads to the removal of PTMs of type P.

The wave function is of form (\ref{ansatz1}) with the local  wave functions  given by
    \begin{equation}
|\Psi_i(t)\rangle 
 = \sum_{p=0}^{S^P} \sum_{m=0}^{S^M}  C_{i,p,m}(t) |p\rangle |m\rangle
 \hspace{1cm}  \langle \Phi_i| = \sum_{p=0}^{S^P} \sum_{m=0}^{S^M} \langle p|\langle m | e^{\phi_{i,p,m}} ,
  \end{equation}
  where the normalization condition $\sum_{p=0}^{S^P} \sum_{m=0}^{S^M} C_{i,p,m}=1$ applies.
  The local  operators $\mathcal{R}_i$, $\mathcal{L}_i$, $\mathcal{N}_i$, $\mathcal{M}_i$, $\mathcal{I}_i$ are defined as
  in section~\ref{mapping}, and we denote the identity  operator by $\mathcal{E}_i ^X$ (unity matrix of size $S^X$). 
Using this notation,  the ``non-hermitian Hamiltonian" of the system is given by $\mathcal{H}=\sum_{i=1}^N \mathcal{H}_i$, where 
 \begin{eqnarray}
\mathcal{H}_i &=& [\lambda^P+ \alpha^P   ( \mathcal{N}^P_{i-1} + \mathcal{N}^P_{i+1} -2\mathcal{N}^P_i) +\tilde{\alpha}^P \mathcal{N}_i^P ]
  (\mathcal{R}^P_i   - \mathcal{I}^P_i) \mathcal{E}_i^M \nonumber \\
&&  +\mathcal{E}_i^P [\lambda^M +\alpha^M  ( \mathcal{N}^M_{i-1} + \mathcal{N}^M_{i+1} -2\mathcal{N}^M_i  ) +\tilde{\alpha}^M \mathcal{N}_i^M      ]
   (\mathcal{R}^M_i  - \mathcal{I}^M_i) \nonumber \\
&&+ [\mu^P+\beta^P  (\mathcal{M}^P_{i-1}  + \mathcal{M}^P_{i+1} -2\mathcal{M}^P_i)+\tilde{\beta}^P \mathcal{M}_i^P]
    (\mathcal{L}^P_i - \mathcal{N}^P_i) \mathcal{E}_i^M \nonumber \\ 
  &&  + \mathcal{E}_i^P[\mu^M+\beta^M    (\mathcal{M}^M_{i-1}  + \mathcal{M}^M_{i+1} -2\mathcal{M}^M_i)+\tilde{\beta}^M \mathcal{M}_i^M]
 (\mathcal{L}^M_i - \mathcal{N}^M_i) \nonumber \\
 &&+ \beta^{P\to M} ( \mathcal{L}^M_i - \mathcal{N}^M_i)   \mathcal{N}_i^P 
 +\beta^{M\to P}  ( \mathcal{L}^P_i - \mathcal{N}^P_i)   \mathcal{N}_i^M.
   \label{QM_H_two}
   \end{eqnarray}
 The master equation in quantum variational formulation becomes
   \begin{equation}
   \left (  \left \langle \frac{\partial \Phi}{\partial \phi_{i,p',m'}} \right | \left .  \frac{\partial \Psi}{\partial C_{i,p,m}} \right \rangle\frac{dC_{i,p,m}}{dt}
       -    \left \langle \frac{\partial \Phi}{\partial \phi_{i,p',m'}} \left  | \mathcal{H} \right  |  \Psi \right \rangle \right )_{\phi_{i,p',m'=0} }= 0.
       \label{variational2}
     \end{equation}
    Evaluating (\ref{variational2}) yields a system of nonlinear difference equations
for the probabilities $C_{i,p,m}$ that the nucleosome at site $i$ has $p$ modifications of type P and $m$ modifications of type $M$,
       \begin{eqnarray}
       \frac{dC_{i,p,m}}{dt} &=& -(\lambda^P +\alpha^P F^{\nabla_P}_i +\tilde{\alpha}^P \langle n_i^P\rangle )  
       (C_{i,p,m}-C_{i,p-1,m})  -(\lambda^M +\alpha^M F^{\nabla_M}_i+\tilde{\alpha}^M \langle n_i^M\rangle )   (C_{i,p,m}-C_{i,p,m-1}) 
\nonumber \\
&&-(\mu^P +\beta^P G^{\nabla_P}_i    +\tilde{\beta}^P \langle m_i^P\rangle            )(pC_{i,p,m}-(p+1)C_{i,p+1,m}) 
 -(\mu^M +\beta^M G^{\nabla_M}_i + \tilde{\beta}^M \langle m_i^M \rangle )    (mC_{i,p,m}-(m+1)C_{i,p,m+1})    \nonumber \\
 && -\beta^{M\to P} \langle n_i^P \rangle  (pC_{i,p,m}-(p+1)C_{i,p+1,m}) 
 -\beta^{P\to M} \langle n^M \rangle  (mC_{i,p,m}-(m+1)C_{i,p,m+1}) .
\label{PDE_system2}
\end{eqnarray}
Here $p=0,1,...,S^P$, $m=0,1,...,S^M$, and 
\begin{displaymath}
\begin{array}{llllll}
F^{\nabla_X}_i = \langle n^X_{i-1}\rangle-2\langle n^X_i \rangle + \langle n^X_{i+1}\rangle &{\rm if } \;\;\;1< i < N,  \hspace{1cm} &  F^{\nabla_X}_1 =-2\langle n^X_1\rangle +  \langle n^X_{2}\rangle  \hspace{1cm} &
F^{\nabla_X}_N =  \langle n^X_{N-1}\rangle -2\langle n^X_N \rangle ,
\\[2mm]
G^{\nabla_X}_i=\langle m^X_{i-1}\rangle-2\langle m^X_i\rangle + \langle m^X_{i+1}\rangle & {\rm if} \;\;\; 1< i < N,
 \hspace{1cm} &  
 G^{\nabla_X}_1 =  -2\langle m^X_1\rangle +\langle m^X_{2}\rangle
\hspace{1cm} &
G^{\nabla_X}_N =  \langle m^X_{N-1}\rangle -2\langle m^X_N\rangle , &
\end{array}
\end{displaymath}
where $X\in \{P,M\}$,
and
\begin{displaymath}
\begin{array}{rclrcl}
\langle n^P_{i}\rangle &=& \sum_{p=0}^{S^P}\sum_{m=0}^{S^M}  p C_{i,p,m} ,&\hspace{2cm}
\langle m^P_{i}\rangle& =& \sum_{p=0}^{S^P}\sum_{m=0}^{S^M}  (S^P-p) C_{i,p,m} = S^P - \langle n_i^P\rangle ,\\[2mm]
\langle n^M_{i}\rangle &=& \sum_{p=0}^{S^P}\sum_{m=0}^{S^M}  m C_{i,p,m} ,
,&\hspace{2cm}
\langle m^M_{i}\rangle& =& \sum_{p=0}^{S^P}\sum_{m=0}^{S^M}  (S^M-m) C_{i,p,m} = S^M  -\langle n_i^M\rangle.
\end{array}
\end{displaymath}
In Eqs.(\ref{PDE_system2}), 
corrections for left-hand-side values of $p=0$, $m=0$, $p=S$, and $m=S$ have to be taken into account, similarly as
 in the first and third equation of (\ref{PDE_system}). 

We consider the case of  inhibition in only one direction by setting $\beta^{M\to P}=0$ and 
 varying $\beta^{P\to M}$,  as is the 
 case in the example mentioned above where Ser10 phosphorylation inhibits
 Lys9 methylation. We evaluate steady states as explained in section (\ref{ODE}).
 For parameter choices of  $S^P=S^M=2$, $\lambda^P=\mu^P=\lambda^M=\mu^M=1$, $\beta^P=\beta^M=3$, $\alpha^P=\alpha^M=4.5$,
  $\tilde{\alpha}^P=4\alpha^P$, $\tilde{\alpha}^M=4\alpha^M$,   $\tilde{\beta}^P=4\beta^P$ and $\tilde{\beta}^M=4\beta^M$, 
  we observe that for small $\beta^{P\to M}$, all four stable steady states (as listed above) exist, as shown in Fig.~\ref{Two_type_bifur}.
  In an intermediate parameter regime  only three stable steady states persist: 00, P0 and 0M, using the notation introduced above (Fig.~\ref{Two_type_bifur}). For large enough $\beta^{P\to M}$, only steady states 00 and P0 remain.
  This means that inhibitory interactions of two types of PTMs  in only {\it one} direction are sufficient to 
 obtain a parameter regime where steady states have either a high number of PTMs P or M, or neither, but not both.
 An analysis of traveling wave solutions of equations (\ref{PDE_system2}) similar to the one in section~\ref{SpatialDependance} 
  applies in this case.


\section{Conclusions and Outlook}

The main results of this paper are as follows. 
 We offer a robust method to obtain nonlinear partial differential equations describing the effective dynamics 
   of histones. The method proceeds by mapping the system  onto a
    quantum spin system whose dynamics is generated by a  non-hermitian Hamiltonian.
     A feedback mechanism due to  diffusion of enzymes along nucleosomes gives rise to multiple stable histone states.
     We study a number of novel aspects in  histone systems that have not been reported before and are of biological relevance.
   We show that explicit cooperativity is not required to obtain multiple stable steady states  as long as the number of PTMs is larger
   or equal to two, and we study the effects of varying the number of PTMs that are regulated by a particular set of enzymes.
    We also study the effect of spatially heterogeneous enzymatic on the histone state, and we 
   apply our approach to a system of several correlated PTMs. 

 Our approach can easily be generalized to higher spatial dimensions and more complicated network topologies.
Processes other than the ones considered in this work could be included into the master equation and 
other biological systems might be studied.
 In the context of post-translational histone modifications, it might be of interest to consider 
   more complex and more realistic systems.
For example, the particular structure of the core histones might be taken into account
 i.e., the exact arrangement of the different modifications on the different core histones.
Feedback processes among different types of post-translational modifications  might be considered, as well as  feedback loops that 
arise due to interactions between the histones and the DNA in the chromatin.
It also remains an open question to study the existence and stability of traveling wave solutions in the nonlinear
 reaction-diffusion equations
 that arise in our model from a mathematically rigorous point of view.

{\it Acknowledgement.--} 
We thank an anonymous referee for very helpful comments and suggestions that improved the
presentation of the results in the paper.



\begin{thebibliography}{10}
\bibitem{Mean_field}
 In this work, we denote as mean-field description an effective  continuum deterministic description (i.e., ordinary or partial differential 
equations). We obtain our 
mean-field Eqs. (11) by rescaling parameters in the difference equations (10) as explained, a procedure
 generally denoted as mean-field in the applied mathematics community.
\bibitem{creation_annihilation}
We note that the operators $\mathcal{R}$ and $\mathcal{L}$ do not correspond to bosonic creation and annihilation operators
(bosonic commutation relations are not obeyed), but are the suitable operators for our model where $S$ 
takes finite values.
\end{thebibliography}
\end{document}